\documentclass[twocolumn,showpacs,preprintnumbers,amsmath,amssymb]{revtex4}

\usepackage{graphicx}
\usepackage{dcolumn}
\usepackage{bm}

\begin{document}

\title{Phase Diagram and Incommensurate Phases in Undoped Manganites}
\author{J. Salafranca and L. Brey}

\affiliation{\centerline { Instituto de Ciencia de Materiales de Madrid
(CSIC),~Cantoblanco,~28049~Madrid,~Spain.}}
  \

\date{\today}

\begin{abstract}
We study the existence of  incommensurate phases  in the phase diagram of the two orbital double exchange model
coupled with Jahn-Teller phonons and with superexchange interactions. In  agreement with experimental results,
we find that undoped manganites $RMnO_3$ ($R$ being some rare earth element) show  temperature induced
commensurate-incommensurate phase transitions. In the incommensurate phase the  magnetic wave vector varies with
temperature. The incommensurate phase arises from the competition between the short range antiferromagnetic
superexchange interaction and the long range ferromagnetic double exchange interaction.
\end{abstract}

\maketitle
\section{introduction}

 Perovskites of manganese of formula (R$_{1-x}$A$ _x$)MnO$_3$ where R denotes rare earth ions(R= La, Pr, Nd
...) and A is a divalent alkaline ion (A=Ca,Sr...) have attracted a great interest because they show a
remarkable colossal magnetoresistance effect at doping $x$ near one third\cite{Tokura,f95-2}. In these oxides
$x$ corresponds to the concentration of holes moving in the $e_g$ orbital bands of the Mn ions that ideally form
a cubic structure. From the basic point of view these materials are a challenge for both theoreticians and
experimentalists as they show a very rich phase diagram\cite{Dagottobook}. As function of temperature and hole
doping, these systems present orbital, charge or spin order, and  in a large portion of the phase diagram these
orders coexist\cite{Goodenough, Radaelli_1997,Tomioka_2002,Rivadulla_2002,Aladine_2002,Greiner_2004, Solovyev,
 brink,Ferrari,Calderon2,brey_2004,brey_2005}. Nanophase separation near $x$=1/3\cite{Uehara_1999} and commensurate incommensurate
transition near half doping seem to occur in colossal magnetoresistance
manganites\cite{chen1,Loudon05,milward05,brey_2005b}.

Many properties of  manganites  depend on the competition between the kinetic energy tending to delocalize the
carriers and localization effects such as the Jahn-Teller (JT) coupling and the antiferromagnetic (AFM) coupling
between the Mn core spins. Therefore the properties of manganites at intermediate doping can be described within
a band structure picture, where the itinerant $e_g$ carriers  have a strong ferromagnetic interaction with the
core $t_{2g}$ Mn spins, and are coupled with the Jahn-Teller distortions of the oxygen octahedra surrounding the
Mn ions\cite{Dagottobook,brey_2005}. However, the parent compounds, RMnO$_3$, are always insulator and their
physical properties were typically described in the picture of strongly correlated Mott localized
$d$-electrons\cite{Mizokawa_1996,Maezono_1998,Brink_1999,Feiner_1999,Capone_2000,Ahn_2001,Okamoto_2002}. However
some spin and orbital ordering found experimentally in undoped
materials\cite{Munoz_2001,Kimura_2003,Kajimoto_2004} can not be easily described in this picture of strongly
localized electrons.

In ref.\cite{Kimura_2003} the authors examine the magnetic and orbital order in a series of RMnO$_3$ as a
function of the ionic radius ($r_R$) of the rare earth ion  R. For small ionic radius the manganites have a
antiferromagnetic spin order of type $A$ coexisting with a ($\pi, \pi, 0$) orbital ordering, whereas for larger
values of $r_R$ the magnetic order is of type $E$. In the $A$ phase, a  Mn spin  is   ferromagnetically  coupled
with the Mn spins located in the same plane ($x-y$), and antiferromagnetically with the Mn spins belonging to
different planes. In the $E$ phase the $x-y$ layers are  antiferromagnetically coupled, but  the magnetic
structure within the planes is that of ferromagnetic zigzag chains coupled antiferromagnetically. The horizontal
($x$) and vertical ($y$) steps of the zigzag chains contain two Mn ions. For values of $r_R$ close to  the
critical value where the magnetic order changes from $A$-type to $E$-type, the manganites develop  different
magnetic incommensurate phases when increasing temperature.
%

It has been suggested recently\cite{hotta_2003a,hotta_2003b,Efremov_2004}, that the complete nesting between the
two $e_g$ bands that occurs in the $A$ structure produces a spin-orbital ordering and opens a gap in the energy
spectrum of undoped RMnO$_3$. Quoting ref.{\cite{Efremov_2004}, we do not claim that the real RMnO$_3$ systems
can be fully described by a weak coupling approach as correlation effects can be important, although  a
treatment based in band structure calculation may be very useful to understand some properties of these
materials. In particular, in ref.\cite{hotta_2003a}, using a two orbital double exchange model, it  was obtained
that the experimental observed $E$-phase exits in a wide region of parameter space, and it is adjacent to
the $A$-type phase.

One of the issues that remains to be understood is the microscopic origin of the commensurate phases appearing
near  the  $A$-type to $E$-type magnetic  transition. The aim of this work is to explain these phases using a
realistic microscopic model.  The Hamiltonian we study  describes electrons moving in two $e_g$ bands, that are
ferromagnetically strongly coupled  to the Mn core spins as well to the Jahn-Teller phonons. In addition, we
also consider a direct superexchange interaction  between the core Mn spins. Starting from this Hamiltonian we
derive a functional that describes a temperature induced commensurate-incommensurate transition similar to that
observed experimentally.

The main result of this work is that near  the $A$ to  $E$  phase transition, the competition between the
nearest neighbor antiferromagnetic superexchange interaction and the double exchange induced long range
ferromagnetic interaction, results in the appearance of incommensurate phases. These phases consist of a
periodic array of domain walls.

The rest of the paper is organized as follows. In Sec.II we describe the microscopic model and we present the
zero temperature phase diagram. In Sec.III we outline the method for obtaining the critical temperatures and we
present the phase diagram composed of the different uniform phases. In Sec.IV we develop the functional for
describing spatially modulated phases. Also in Sec.IV we study how  the phase diagram of manganites at $x$=1 is
altered when soliton incommensurate phases are taken into account. We finish in Section V with a brief summary.

\section{Microscopic  Hamiltonian and zero temperature phase diagram.}
We are interested in the transition between the $A$ and $E$ phases. In these phases the $x-y$ planes are coupled
antiferromagnetically and therefore we can analyze the properties of these states and the transition between
them by studying a Hamiltonian which describes electrons moving in the $x-y$ plane. The Coulomb interaction
between electrons prevents double occupancy and aligns the spins of the $d$ orbitals. The  crystal field splits
the Mn $d$ levels into an occupied  $t_{2g}$ triplet and a doublet of $e_g$ symmetry where $1-x$ electrons per
Mn have to accommodate. The Hund's coupling between the spins of the $e_g$ electrons  and each core spin is much
larger than any other energy in the system, and each electron spin is forced to align locally with the core spin
texture. Then the $e_g$ electrons can be treated as spinless particles and the  hopping amplitude between two Mn
ions is modulated by the spin reduction factor,
\begin{equation}
f_{12}= \cos\frac{\vartheta _1}{2}\cos\frac{\vartheta _2}{2} + e ^{i ( \phi _1 - \phi _2)} \sin\frac{\vartheta
_1}{2}\sin\frac{\vartheta _2}{2} \label{SRF}
\end{equation} where
$\{\vartheta _i, \phi _i \}$ are the Euler angles of the, assumed classical, Mn core spins $\{\textbf{S} _i \}$.
 This is the so called  double exchange (DE) model\cite{Zener,Anderson,DeGennes}.

We study a  double exchange  model coupled to Jahn-Teller (JT) phonons. We also include the antiferromagnetic
coupling between the Mn core spins $J_{AF}$.
\begin{eqnarray}\label{Hamiltonian}
 H &  = &  -\sum_{i,j,a,a'} f _{i,j} t _{a,a'} ^{u} C ^+ _{i,a} C
_{j,a'}
\nonumber \\
 &+& J_{AF} \sum _{<i,j>}
\textbf{S} _i \textbf{S} _j
 + \frac{1}{2} \sum _i \left ( \beta Q _{1i} ^2+ Q
^2_{2i} + Q _{3i} ^2 \right )
\nonumber \\
 & + & \lambda \sum _{i} \left ( Q _{1i} \rho _i + Q _{2i} \tau _{xi}
+ Q _{3i} \tau _{zi} \right ) \, \, \, ,
\end{eqnarray}
here $C^+ _{i,a}$ creates an electron in  the Mn ion located at site $i$, in the $e_g$ orbital $a$ ($a$=1,2 with
1=$|x^2-y^2>$ and 2=$|3z^2-r^2>$). The hopping amplitude $t_{aa\prime} $ is finite for next neighbors  Mn and
depends both on the type of orbital involved and on the direction $u$ between sites $i$ and $j$
($t_{1,1}^{x(y)}=\pm \sqrt{3} t_{1,2}^{x(y)} =\pm \sqrt{3} t_{2,1}^{x(y)}=3t_{2,2}^{x(y)}=t)$\cite{Dagottobook}.
$t$ is taken as the energy unit.  The forth term couples the $e_g$ electrons with the three active MnO$_6$
octahedra distortions: the breathing mode $Q_{1i}$, and the JT modes $Q_{2i}$ and $Q_{3i}$ that have symmetry
$x^2$-$y^2$ and $3z^2$-$r^2$ respectively. $Q_{1i}$ couples with the charge at site $i$,  $\rho _i = \sum _a C^+
_{i,a} C _{i,a}$ whereas $Q_{2i}$ and $Q_{3i}$ couple with the $x$ and $z$ orbital pseudospin, $\tau _{xi}= C^+
_{i1}C_{i2} + C^+ _{i2} C_{i1}$ and $\tau _{zi}= C^+ _{i1}C_{i1} - C^+ _{i2} C_{i2}$, respectively.  The third
term is the elastic energy of the octahedra distortions, being $\beta \geq$2 the spring constant ratio for
breathing and JT-modes\cite{Aliaga}. In the perovskite structures the oxygens are shared by neighboring MnO$_6$
octahedra and the $Q$'s distortions are not independent, cooperative effects being  very important\cite{JAV}. In
order to consider these collective effects, we consider the position of the oxygen atoms as the independent
variables of the JT distortions.

For a given value of the parameters $ \lambda$ and  $J_{AF}$, and a texture of core spins $\{ \textbf{S}_i \}$,
we solve self-consistently the mean field version of Hamiltonian (\ref{Hamiltonian}) and obtain the energy, the
local charges $\{\rho_i\}$, the orbital pseudospin order $\{\tau _{xi}, \tau_{zi} \}$ and the oxygen octahedra
distortions $Q_{\alpha,i}$. These quantities are better described by their Fourier transforms, that are
represented by the same symbol with a hat: $\hat{\rho} (\bf{G})$, $\hat{Q _1} (\bf{G})$, ...

\begin{figure}
  \includegraphics[clip,width=8cm]{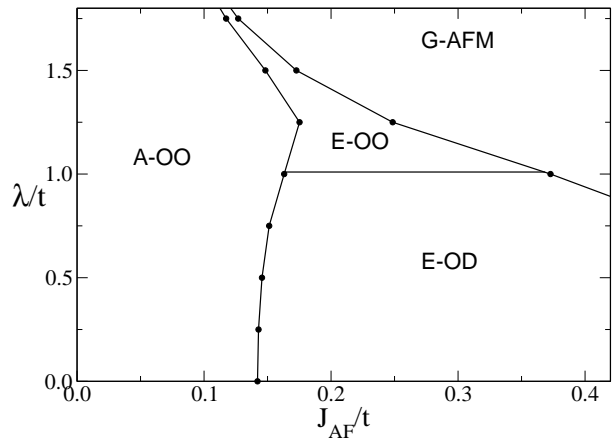}
  \caption{Zero temperature phase diagram for $x$=1 for the two-dimensional  DE
  two orbital model with cooperative Jahn Teller phonons. The
  symbols
    $OO$  and $OD$ stand for orbital ordered and disordered respectively. $A$, $E$ and $G$  name
     the different magnetic order defined in the text.}
\end{figure}

In Fig.1 we present the phase diagram obtained by solving self-consistently Eq.\ref{Hamiltonian} for the parent
compound RMnO$_3$.  For the range of parameters studied, we do not find any  solution showing charge modulation.
In all the phases there  is an electron located on each Mn ion, therefore in our model any gap in the energy
spectrum is due to the spatial modulation of any other physical quantity.

For small values of $J_{AF}$ the ground state is ferromagnetic, $A$-order. In absence of Jahn-Teller coupling
this phase is metallic, however, for $\lambda \ne$0, and due to the perfect nesting between the $e_g$ bands, the
$A$ phase  develops a gap at the Fermi energy. The Jahn-Teller coupling produces and orbital  order
characterized by a finite Fourier component of the $x$-component of the pseudospin $\hat{\tau
_x}(\pi,\pi)=\hat{\tau _x}(-\pi,-\pi)\neq 0$, see Fig.2a.  The orbital order (OO) is produced by an ordered
distribution of the oxygen octahedra distortions $\hat{Q _2}(\pi,\pi)=\hat{Q _2}(-\pi,-\pi)\neq 0$, that depends
on the value of $\lambda$. The amplitude of the distortions are modulated in order to minimize the elastic
energy of the cooperative Jahn-Teller distortions, and the signs arise from cooperative effects. In this phase
the $(\pi,\pi)$ orbital modulation opens a gap at the Fermi energy and the $x$=1 manganite is an insulator,
being the energy gap proportional to the value of the Jahn-Teller coupling.

For large value of $J_{AF}$ and $\lambda$ the system presents a $G$-type  antiferromagnetic  ground state and an
orbital order characterized by a Fourier component of the pseudospin $\hat{\tau _x}(\pi,\pi)=\hat{\tau
_x}(-\pi,-\pi)\neq 0$. Each Mn ion is coupled antiferromagnetically with its next neighbors and the double
exchange mechanism precludes the motion of the carriers, being this phase an insulator. The minimal value of
$J_{AF}$ for the occurrence of this phase depends on $\lambda$, but in general is very large, so that this phase
is rather unlike to occur in manganites.

\begin{figure}
  \includegraphics[clip,width=9cm]{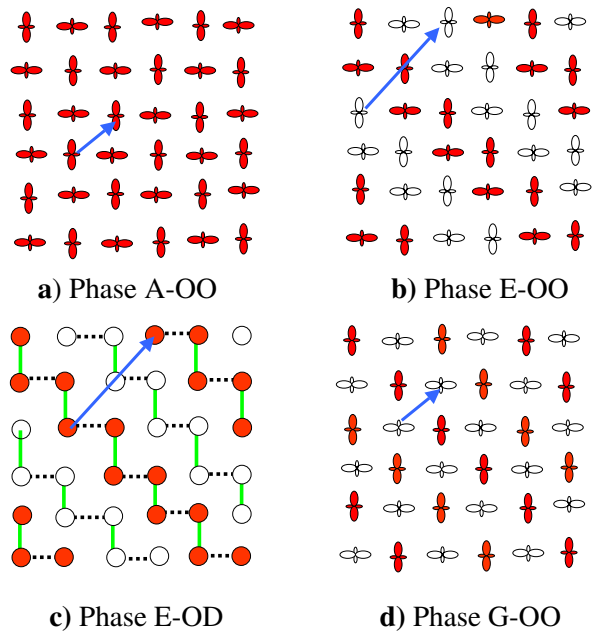}
  \caption{(Color online) Orbital and spin order of $x$=1 manganites in the $x$-$y$ plane.
  Elongated orbitals along the $x$ ($y$) directions represent $d_{3x^2-r^2}$ ($d_{3y^2-r^2}$) orbitals.
  Circles represent the Mn ions in a orbital disordered phase.
{\bf a}) Orbital order presents in the ferromagmetic $A$-$OO$ phase. {\bf b}) Same than {\bf a} but for the $E$-$OO$
phase.  {\bf c}) Spin order in the $E$-$OD$ phase. The solid and the dashed lines joining the Mn ions indicate the
modulation of the electronic coherence along the zigzag chains. {\bf d}) Same than {\bf a} but for the $G$-$OO$
phase. In all the phases there is not modulation of the electric charge and there is an electron located at each
Mn ion. The vectors in the different schemes, represent the spatial periodicity in the different phases, $(a,a)$
in the $A$-$OO$ and $G$-$OO$ phases and $(2a,2a)$ in the $E$-$OO$ and $E$-$OD$ phases. In all the figures open and close symbols
represent up and down spins.}
 \label{Fig2}
\end{figure}

For intermediates values of $J_{AF}$ the system develops a magnetic order of $E$-type; the $E$ phase consists of
ferromagnetic zigzag chains coupled antiferromagnetically. The horizontal and vertical steps of the chain contain
two Mn ions. For large enough  values of the Jahn-Teller coupling the $E$ magnetic order coexists with an
orbital order similar to the one occurring  in the FM and $G$ phases; this order opens a gap at the Fermi energy. In
the $E$-$OO$ phase the magnetic order is characterized by a periodicity $(2a,2a)$, being $a$ the lattice parameter
of the square lattice. For small values of $\lambda$ the $E$ phase does not present orbital order, although it
has a gap at the Fermi energy. In the orbital disorder ($OD$) $E$-phase, the dispersion energy for the $e_g$
electrons along the FM zigzag chain is given by\cite{hotta_2003a}, $\varepsilon _k = (2/3) (\pm \cos k \pm
\sqrt{ {\cos} ^2{k} + 3}) $, indicating the existence of  a large band gap at occupancies corresponding to
$x$=1. The physical origin of this gap is the dependence of the tunnelling probability  on the spatial
direction, $t_{\mu,\nu} ^x = - t_{\mu,\nu} ^y $ for $\mu \ne \nu$. It produces a periodicity in the hopping amplitude along the zigzag chain, leading to a periodic potential for the $e_g$ electrons. It is
important to note that, contrary to the $x$=1/2 case\cite{brey_2004,brey_2005}, this modulation in the hopping
amplitude does not produce an orbital order. \emph{The $E$-$OD$ phase is stable due to the spatial modulation of
the coherence between next neighbors Mn ions along the zigzag chain}, $<C^+ _{i,\mu} C_{i+1,\nu}> = - <C^+
_{i+1,\mu} C_{i+2,\nu}>$ for $\mu \ne \nu$. The phases $E$-$OO$ and $E$-$OD$ have different symmetry and therefore
the transition between this phases that occurs at finite values of $\lambda$ is a discontinuous transition.

\section{Finite Temperature Magnetic  Phase Diagram.}
The simplest way of obtaining information on the phase diagram corresponding to  a given microscopic Hamiltonian
is by means of the mean field approximation. This approximation is insufficient  for describing  second order
transitions, but it is successful in describing the phases away from the transition and in predicting  the topology
of the phase diagram. In this approach we compute the magnetic critical temperature of the different phases.
\subsection{Ferromagnetic  (A) phase}
In this phase all the Mn spins point, on average, in a particular direction, and there is a finite relative
magnetization $< \!  \! m  \! \! >$. Using a virtual crystal approximation, we consider a unique value for the
spin reduction factor $f_{i,j}$ that corresponds to its expectation value\cite{Arovas-1999,brey_2005},
\begin{equation} f_{ij}\simeq <  \! \! \sqrt{\frac{1
\! + \! \cos{\theta_{ij}}}{2}} \!
> \! \simeq  \! \sqrt{\frac{1 \! + \! < \!
\cos{\theta_{ij}} \! >}{2}} \! = \! \sqrt{\frac{1 \! +  \! \! < \! \! m \! \!
>^2}{2}}. \label{srf_ap}
\end{equation}
A reduction of $<\! \! m \! \! >$ produces a decrease  in $f_{i,j}$ and therefore in the kinetic energy. In this
way the  importance of the Jahn-Teller coupling increases as the temperature decreases. The internal energy per
$Mn$ ion of this phase can be written as
\begin{equation}
E^{A}= \varepsilon _{A} (\lambda,< \! \!  m \!  \!>) + 2 J_{AF} < \! \! m \! \!> ^2 \, \, \, .
\label{energy_FM_CO}
\end{equation}
where the electronic energy per Mn ion, $\varepsilon _{A} (\lambda,< \! \! m \! \!>)$, depends in a complicated
way on $\lambda$ and $< \! \! m \! \! >$ and has to be obtained numerically by solving Eq.(\ref{Hamiltonian}).

In order to describe thermal effects it is necessary to compute the free energy. As the entropy of the carriers
is very small\cite{brey_2004} we only  include the entropy of the classical Mn spins. We use a mean field
approximation that neglects spatial correlations and  assume for each individual spin a statistical distribution
corresponding to an effective magnetic field\cite{DeGennes,brey_2004}. In this molecular field approximation the
entropy of the Mn spins takes the form,
\begin{equation}
S ( <\! m \! >) =\frac{\log 2}{2}-\frac{3}{2} <\! \! m  \! \!> ^2- \frac{9}{20} <\! \! m \! \!> ^4 + ....
\label{expansion}
\end{equation}
Using this expression for the Mn spins entropy the total free energy of the system for small values of $< \! \!
m\! \!>$ takes the form
\begin{equation}
F( <\! \!  m \! \! > ) =  E^{A}-  \! T S ( <\! \! m  \! \!>) \nonumber \\
\end{equation}
And the  Curie temperature of the $A$ phase is,
\begin{equation}
T_C = - \frac{2}{3} \left . \frac{\partial \varepsilon^{A} (\lambda, <\! \!m\! \!>)}{\partial <\! \!  m \! \!
> ^2}    \right  | _{< \! \! m \! \! > = 0}  - \frac{4}{3} J_{AF}\, \, \, .
\label{tcfm}
\end{equation}
For finite $\lambda$ the derivative has to be calculated numerically. From higher derivatives of the internal
energy with respect to the magnetization we obtain that the transition is second order. In Eq.\ref{tcfm} we
notice that for a given value of the Jahn-Teller coupling the Curie temperature decreases linearly with the
superexchange antiferromagnetic coupling $J_{AF}$.

\subsection{Antiferromagnetic $E$ phase.}
The magnetization of the $E$ phase is described by the relative amount of saturation in each zigzag chain $<\!
\! m_s \! \!>$. In the virtual crystal approximation fluctuations are neglected and the  hopping is modulated
by the spin reduction factor that is different along the zigzag FM chain, $f ^ {FM}$ than between the AFM coupled
chains, $f ^ {AF}$\cite{DeGennes,brey_2005},
\begin{eqnarray}
  f ^ {FM}(<\! \!m_S \!\! >) &=& \sqrt{\frac{1+<\! \!m_S \!\! >^2}{2}} \nonumber \\
  f ^ {AF}(<\! \!m_S \! \!>) &=& \sqrt{\frac{1-< \! \!m_S \! \!>^2}{2}}\, \, \, .
\label{srf_CE}
\end{eqnarray}
The internal energy of this phase depends on  $\lambda$, and $<\!\!  m_S \! \! >$, and can be written as,
\begin{equation}
E^{E}= \varepsilon _{E} (\lambda,< \!\! m_S \! \!>)  \, \, \, . \label{energy_CE_CO}
\end{equation}
As each $Mn$ spin core is surrounded by two $Mn$ spins coupled FM and other two coupled AFM, the superexchange
energy is zero.

In order to compute the Neel temperature of the $E$ phases, we introduce an effective field for each spin
sublattice. Taking into account that both, the magnetization and the effective magnetic field, have different
sign in each sublattice, we end up with the same expression for the entropy than in the FM phase, but just changing
$<\! \!m \!\! >$ by $<\! \! m  _ S \! \!  >$\cite{brey_2005}. With this the free energy takes the form,
\begin{eqnarray}
F(<\! \!m _ S \!\! >,\lambda) & = & E^{E} - T S ( <\! \! m_S \!\!
>)
\nonumber \\
& \simeq  & F(0,\lambda)+<\!\!m  _ S \!\!> ^2 \left ( \frac{3}{2} T + a \right ) \nonumber
\\ &+ &<\!\!m  _ S \!\!> ^4 \left (\frac{9}{20} T + b \right ) + \nonumber
\\ &+ &<\!\!m  _ S \!\!> ^6 \left (\frac{99}{350} T + c \right ) +
 ...
\end{eqnarray}
with
\begin{equation}
E^{E} \thickapprox cte+ a <\!\! m _S \! \!> ^2 + b <\!\! m _S \!\!
> ^4+ c <\! \!m _S \!\!
> ^6+  ....
\end{equation}
Due to the symmetry of the $E$ phase the coefficient $a$ is zero,
and the Neel temperature depends on the coefficients $b$ and $c$.
Numerically, the coefficient $b$ is negative and the transition
from the $E$ to the  paramagnetic (PM) phase is a first order
phase transition.

It is interesting to analyze the origin  of the negative sign of the quartic term.
In the $OD$-case  the Jahn-Teller coupling is not large enough to produce orbital order. In this situation the
electronic energy is just kinetic energy. Therefore near  $<\!\!m _ S \!\!>$=0  we would expect that the
electronic energy could  be obtained perturbatively from the paramagnetic energy as, $E^{E}_{\lambda=0}
\thickapprox \frac {1}{\sqrt{2}} \left ( f ^{FM} (<\! \!m_S \!\! >) + f^{AF} (<\!\! m_S \!\!
>)\right ) \, \varepsilon _{E} ^0 $.  Here $\varepsilon _E ^0 \equiv \varepsilon _{E} (0 ,0)$ is the paramagnetic energy per Mn ion. Expanding the spin reduction factors near $<\! \! m _S  \! \!>$ we
find $E^{E}_{\lambda=0} \thickapprox \left ( 1 - \frac {1 } {8}< \!\! m_S \!\! > ^4 - \frac{5}{128} < \!\! m_S
\!\!
> ^8 -...\right)\varepsilon _E ^0$. As the electronic energy of the paramagnetic phase is negative, the last
expression suggests that the Neel temperature should be zero. However, numerically,  we find a finite Neel
temperature even for  $\lambda$=0. This discrepancy occurs because, as commented above, in the $E$-$OD$ phase the
minimization of the kinetic energy produces a modulation of the electron coherence along the zigzag chain, $<C^+
_{i,\mu} C_{i+1,\nu}> = - <C^+ _{i+1,\mu} C_{i+2,\nu}>$ for $\mu \ne \nu$. We describe this modulation by a
order parameter $\xi$ that represents the $(\frac{\pi} {2}, \frac{\pi}{2})$ Fourier component of the electron
coherence. This order parameter is coupled with the staggered magnetization and the functional describing the
electronic energy has the general form,
\begin{eqnarray}
E^{E}_{\lambda=0} &  \thickapprox &  \frac {1}{\sqrt{2}} \left ( f ^{FM}(< \! \! m_S \!\! >) + f^{AF}(< \! \!m_S
\! \!
>) \right ) \, \varepsilon _{E} ^0 \nonumber \\ & + &  \alpha \xi ^2 + \beta \xi < \! \! m_S \! \!>^2+ ...
\label{acoplo}
\end{eqnarray}
where we have included the elastic energy associated with the electron  coherence and the minimal coupling
between the staggered magnetization and the electron coherence. Minimizing this energy with respect the
coherence parameter $\xi$, we find $\xi= - \frac {\beta} {2 \alpha} < \! m_S \! > ^2$. Introducing this value
in the expression of the electronic energy,  Eq.(\ref{acoplo}), we obtain,
\begin{equation}
E^{E}_{\lambda=0} \thickapprox \left ( 1 - \frac {1 } {8}< \! m_S \! > ^4 \right)\varepsilon _E ^0 - \frac{\beta
^2} { 4 \alpha} < \! m_S \! > ^4 + ...
\end{equation}
and for strong enough coupling between $< \! m_S \! >$ and the orbital coherence, the quartic term is negative
and a finite   Neel temperature is expected. It is therefore the coupling between the electron coherence and the
staggered magnetization the responsible for the occurrence of a finite Neel temperature.

In the $OO$-case there exits a finite orbital order parameter $\hat{\tau _x}(\pi,\pi)$, that is coupled with the
staggered magnetization and it is the responsible of the existence of finite Neel temperature.

\subsection{Temperature-$J_{AF}$ magnetic phase diagram}
\begin{figure}
  \includegraphics[clip,width=8cm]{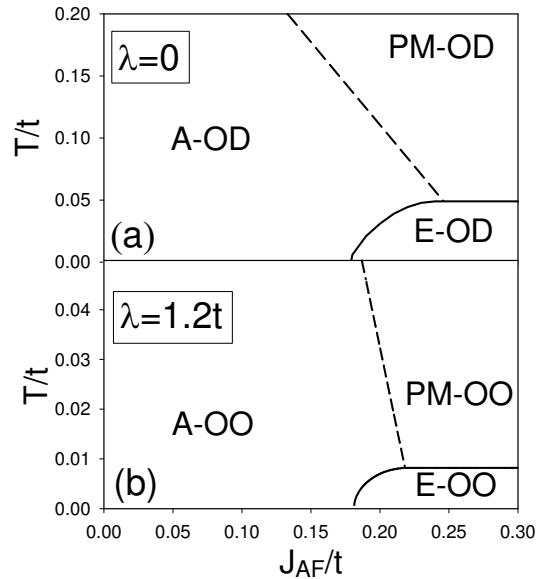}
  \caption{Phase diagrams $T$-$J_{AF}$ for the two dimensional DE two orbital model with cooperative
  Jahn-Teller phonons and $x$=1. In a) we plot the $\lambda =0$ case and in b) the $\lambda=1.2t$ case. Continuous lines represent first order transitions whereas dashed lines indicate
  second order transitions. The abbreviations naming the different phases are explained in the text.}
 \label{Fig3}
\end{figure}
In Fig.3, we plot the T-$J_{AF}$ magnetic phase diagrams for $\lambda$=0  (a) and  $\lambda$=1.2$t$ (b). These
phase diagrams have been obtained by minimizing and comparing the free energy of the $A$, $E$ and paramagnetic
phases. For $\lambda$=0, all the phases  are disordered in the orbital sector, however, for large enough values
of $\lambda$, the $E$ and $A$ phases present orbital order. In the later case, $\lambda$=1.2t, we find that the
critical temperature associated with the orbital order is much larger than the magnetic critical temperatures
and, therefore, in Fig.3b the paramagnetic phase presents orbital order. In any case, it  is important  to note
that, from the magnetic point of view, both phase diagrams are topologically equivalent. At large temperatures
the systems are  always paramagnetic, for small $J_{AF}$ and small temperature the systems present ferromagnetic
order, whereas for small temperature and moderates values of $J_{AF}$ an antiferromagnetic order of type $E$
appears. For very large values of the AFM coupling, not shown in Fig.3,  antiferromagnetic order of type $G$
would appear. The Curie temperature corresponding to the paramagnetic-$A$ phase transition decreases linearly
with $J_{AF}$, Eq.6,  until it reaches the,  $J_{AF}$ independent,  Neel temperature corresponding to the
paramagnetic-$E$ transition. As discussed in the previous subsection the $A$-paramagnetic transition is second
order while, because of the coupling between different order parameters, the $E$-paramagnetic transition is
first order.

The phase diagrams present a Lifshitz point where the uniform ferromagnetic $A$ phase, the modulated ordered $E$
phase and the paramagnetic disordered phase meet. Near the Lifshitz  point there is a range of values of
$J_{AF}$ where, by increasing the temperature, the system undergoes an $E$-$A$ transition followed by an $A$-PM
transition.  The topology of this phase diagram is similar to that of a Ising model with competing interactions.
In that model, near the Lifshitz point, solitons,  spatially modulated phases and commensurate incommensurate
transitions appear  when the temperatures varies\cite{bak_1980}. In the next section we explore the possible
existence of solitons and incommensurate phases in the model described by the Hamiltonian Eq.\ref{Hamiltonian}
at $x$=1 and near the Lifshitz point that appears in the $T$-$J_{AF}$ phase diagram, Fig.\ref{Fig3}.

\section{Soliton theory and spatially modulated phases.}
\subsection{Landau functional}
\begin{figure}
  \includegraphics[clip,width=9cm]{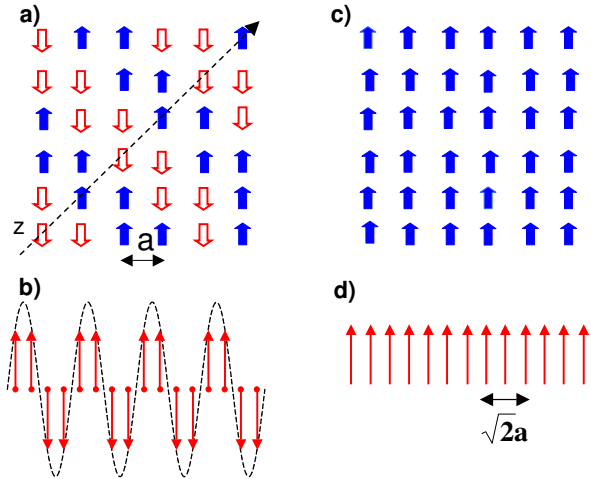}
  \caption{(Color online) Spin order of $x$=1 manganites in the $x$-$y$ plane.
  (a) Corresponds to the $E$ phase whereas (c) represents the $A$ phase.
  The direction perpendicular to the zigzag chains is shown in (a).
  The averaged magnetization along the $z$ direction
  for the $E$ and $A$ phases are plotted in (c) and (d) respectively. }
 \label{Fig4}
\end{figure}
Magnetically, the phases described in the previous sections only
vary along the direction perpendicular to the chains, see
Fig.\ref{Fig2}. In the mean field approximation the $E$ phase is
described by a spin density wave of the form $<\! S \! >=\sqrt{2}
\, m _0 \,  \cos ( q _0 z + \pi /4)$, with $q_0 = \pi / 2$ and
here $z$ is the position of the atoms along the direction
perpendicular to the chains. We are taking the distance between
first neighbors diagonal lines of atoms, $\sqrt{2}a/2$, as the
unit of length. In general the expression
\begin{equation}
<\! S \! >=\sqrt{2} \, m _0 \,  \cos \left ( q _0 z + \frac {\pi}
{4} + \theta (z) \right ) \label{magz}
\end{equation}
describes different spatially modulated magnetic phases. With $\theta(z)$=0 it describes the $E$ phase, whereas
with $\theta (z)$=$-q_0 z$, it represents the average magnetization in the  position independent ferromagnetic
$A$ phase. The case $m_0$=0 corresponds to the paramagnetic case.  In general $\theta (z)$ describes phases
where the average magnetization changes along the direction perpendicular to the chains.

Both the $E$ and $A$ phases are commensurate with the underline lattice, here we are going to study  the
existence of solitons  and incommensurate phases in the system. The solitons are static domain walls between
commensurate domains. In this section we present a formalism which makes contact with phenomenological theories
and provides the basis for the calculation of the nature of the phase diagram at finite temperature.

We want to build a Landau theory functional where the order parameter is the modulation of the average spin
along the diagonal direction. Following reference \cite{McMillan76,McMillan77}, in Eq.\ref{magz} we consider the
amplitude of the magnetization, $m_0$, as constant in the space.

In order to set up  a Landau theory we need to calculate the different contributions to the free energy. For a
magnetization given by Eq.\ref{magz}, and near the order disorder magnetic transitions, the entropic
contribution to the free energy can be estimated as described in Sec.III,
\begin{eqnarray}
-TS \simeq    k _B   T  & &  \! \! \! \! \! \! \! \! \! \! \! \!
\int    \left [    - \log 2 + \frac{3}{2} m _0 ^2 +
\frac{27}{40} m _0 ^4 + \frac{99}{140} m _0 ^6 \right . \nonumber \\
   & - &  \left . (
\frac{9}{40} m _0 ^4 + \frac{297}{700} m _0 ^6  ) \cos { 4 \theta (z)}  \right ] d z \, \, \, \, \, .
\label{entropy}
\end{eqnarray}
The superexchange antiferromagnetic interaction takes the form
\begin{equation}
E_{AF} \simeq 2 J _{AF} \, m _0 ^2 \int \sin { \left ( \nabla \theta (z) \right )}  \, dz \, \, . \label{EAF}
\end{equation}
being $\nabla \theta (z)$ the derivate of $\theta (z)$ with respect $z$. In the previous expression we have
treated the position $z$ as a
 continuous variable and we have discarded second and higher derivates
 of $\theta$ with respect the position.

Concerning the electronic contribution to the internal energy, $E_e$,  we assume  that it is local an can be
written as $E_e$=$\int \! dz  \, {\cal E } (z) $, being ${\cal E}(z) $ the electronic energy density.
We expect that it can be expanded in powers of the order parameter $<\! S \! >$ and its derivates,
\begin{eqnarray}
{\cal E}(z) = {\cal E } ^ 0  & + &  \tilde{a} _l (z) <\! S (z) \! > ^ l +\tilde{ b} _l (z) \, ( \nabla \!\! \!
<\! S (z) \!
>) ^ l \nonumber \\ & + &  \tilde{c}_{l,m}( z) <\! S (z) \! > ^ l  \,  (\nabla \! \! \! <\!  S (z) \! >) ^ m \, \, ... \, \,  \label{expansion}
\end{eqnarray}
Here the sum over repeated indices is assumed, and because the symmetry of the system only even powers of $<\!
\! S \! \!
>$ and $\nabla  \! \! <\! \! S (z) \! \! >$ contribute. The coefficients $\tilde{a}_l$, $\tilde{b}_l$ and $\tilde{c}_{l,m}$ are periodic in
$z$ with the periodicity of the crystal lattice and for a magnetization of the form Eq.\ref{magz}, the density
of electronic energy can be written as,
\begin{eqnarray}
\! \! \! \! \! \! {\cal E} (z) = {\cal E }^ 0  \! \! \!&+&  \! \! \! a _2 m _0 ^2 + a_4 m _0 ^4 + a_6 m _0 ^6 \nonumber \\
\! \! \!& + & \! \! \!( b_4  m_0 ^4 + b _ 6 m_0 ^6 ) \cos {4 \theta (z)} \nonumber \\
\! \! \!& + & \! \! \!( c _2 m _0 ^2 + c_4 m _0 ^4 + c_6 m _0 ^6  ) \, ( \nabla \theta (z) + q_0 ) ^2 .
\label{expansion1} \end{eqnarray} Here we have neglected higher terms in the derivates of the phase $\theta$
and,  as there are first order phase transitions in some part of the phase diagram, we keep terms up to the
sixth power in $m_0$. The second term  in Eq.\ref{expansion1} is the \textit{umklapp }term that favors the
modulated commensurate solutions, $\theta (z) $=0, $\frac {\pi}{2}$, $\pi$, $3\frac {\pi}{2}$, corresponding to
the $E$ phase. The last term is an elastic energy which favors the occurrence of the ferromagnetic $A$-phase,
$\theta (z)$=$-q_0 z$. The competition between the elastic and the  \textit{umklapp } will produce the existence
of solitons and incommensurate phases.

>From the expression of the electronic energy of the $E$-phase as function of the order parameter $m_0$, we
notice  $a_2$=$-c_2 \, q_0 ^2$. Analyzing the dependence of the electronic energy on a   constant phase ($\theta
(r)=\theta_0 $), we find $b_4$=$a_4+c_4 \, q_0 ^4$ and $b_6$=$a_6+c_6 \, q_0 ^4$. In this way only two
subsets of parameters (for instance $b$'s and $c$'s) are independent. Finally, for each value of $\lambda$, we
perform microscopic calculations of the electronic energy of the $A$ and $E$ phases and obtain the numerical
values of the coefficients $b$'s and $c$'s respectively.

Adding the entropy, Eq.\ref{entropy},  the antiferromagnetic energy, Eq.\ref{EAF} and the electronic internal
energy, Eq.\ref{expansion1},  we obtain the following expression for the free energy of the system,
\begin{eqnarray}
\! \! \! \! \! \! \! \! \! \! F \! \! &=  & \! \!  F _0 (T, m_0)  \nonumber \\ & + & \! \! \! C \! \! \int { \!
\! \!  \left [ \frac 1 2 \, \left ( \nabla \theta (z) + q_0 \right ) ^2 + w \left ( 1 + \cos {4 \theta (z) }
\right ) \right ] dz } \label{freeenergy}
\end{eqnarray}
with,
\begin{eqnarray}
F _0 (T, m_0) & =&  \left ( - \log 2 + \frac 3 2 m _0 ^2 + \frac {9}{10} m _0 ^4 + \frac{198}{175} m _0 ^6
\right ) T \nonumber \\ & - &  2 J_{AF} m _ 0 ^2 + \varepsilon _e ^0 - c _2 m _0 ^2 q _ 0 ^2 \, \, \, ,
\label{fcero}
\end{eqnarray}
\begin{equation}
C= \left ( 2 c_2 + 4 \frac {J_{AF} }{ q _0 ^2} \right ) m _0 ^2 + 2 c_4 m _0 ^4 + 2 c_6 m_0^6 \, \, ,  \label{C}
\end{equation}
and
\begin{equation}
w = \frac { \left ( b _4 -\frac {9}{40} \right )  m _0 ^4 + \left ( b_6 - \frac {297}{700} T \right ) m _0 ^6}
{C} \, \, \, . \label{w}
\end{equation}
In the limit  $w \rightarrow 0$, the elastic contribution  is the more important term and the  phase $\theta (z)$
tends to be $\theta (z) = - q _0 \, z$. On the contrary, for large values of $w$, the \textit{umklapp} term is
dominant and $\theta$ wants to get a constant value, $\theta$=0, $\frac {\pi}{2}$, $\pi$, $3\frac{\pi}{2}$. A
transition between the commensurate phase, $\theta = 0 $  and the uniform ferromagnetic phase takes place
because of the competition between these two terms; by tuning the values of $J_{AF}$ and $T$ we are going to see
that a soliton incommensurate phase appears between these two limits.

For a given temperature and a particular value of $J_{AF}$, the constant amplitude $m_0$ and the phase function
$\theta (z)$ that characterize the solution are obtained by minimizing the functional Eq.\ref{freeenergy}. For
each $m_0$ , the phase $\theta (z)$ should satisfy the sine-Gordon equation,
\begin{equation}
\frac {1}{2} \frac { d ^2 \theta}{dz^2} + 4 w \sin{4 \theta} = 0 \, \, \, \, \,  \label{sinegordon}
\end{equation}
which has soliton-like solutions of the form,
\begin{equation}
\theta (z) = \tan ^{-1} \exp {( 4 \sqrt{w} z )} \, \, \, \, .\label{soliton}
\end{equation}
This solution is a domain wall which separates two almost commensurate $z$ regions.

In general the solutions of Eq.\ref{freeenergy} is a soliton lattice formed by a regular array of domain walls, $L$.
At each soliton the phase $\theta$ tumbles $\frac {\pi}{4}$. The deviation of the average wavevector $\bar{q}$
from $q_0$ is inversely proportional to the distance between the domain walls,
\begin{equation}
\bar{q} = \frac { \pi }{2 L } \label{qbar} \, \, \, .
\end{equation}
The value of $\bar{q}$ is proportional to the soliton density and is obtained by minimizing the free energy
following the procedure outlined in references\cite{bak_1980,McMillan76,McMillan77,DeGennes_68}

In the soliton lattice phase the magnetic periodicity along the $z$-direction is characterized by the wavevector
\begin{equation}
q= \frac {\pi }{2} - \frac{\pi}{2L} \, \, \, \, . \label{qreal}
\end{equation}
In the $E$-phase there are not solitons in the system $L$=$\infty$ and the wavevector of the magnetic modulation
is $q$=$\frac {\pi}{2}$. In the continuous approximation  the ferromagnetic $A$-phase corresponds to a extremely
dense lattice soliton, $L$=1. In this limit $\theta$ is too quickly varing, the continuous approximation is not valid and we take the criterium
that for $L\leq 1.1$, the soliton lattice is the ferromagnetic $A$-phase.

\subsection{Results}

\begin{figure}
  \includegraphics[clip,width=8cm]{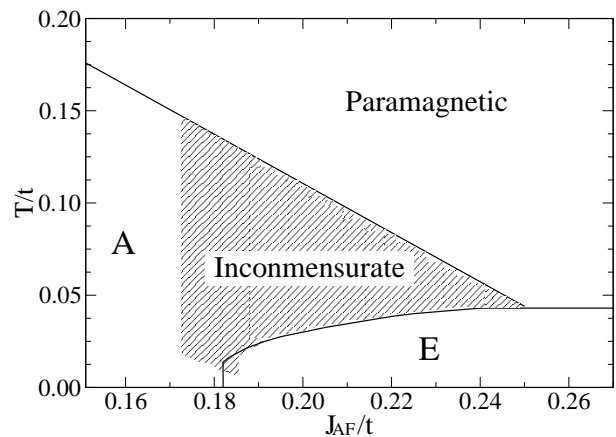}
  \caption{Phase diagram $T-J_{AF}$ as obtained by minimizing the free energy Eq.\ref{freeenergy}.
  The parameters of entering in the free energy are obtained by minimizing the microscopic Hamiltonian Eq.\ref{Hamiltonian}
  for  $\lambda$=0 and $x$=1. Continuous lines represent first order transitions whereas dashed lines indicate
  second order transitions. The shadow region indicates the region where the incommensurate phase exits.}
 \label{Fig5}
\end{figure}

In order to find inhomogeneous phases in manganites at $x$=1, we have minimized the free energy
Eq.\ref{freeenergy} with the coefficients $a$'s, $b$'s and $c$'s obtained from the microscopic model described
in Section II.  The solutions are characterized by the value of the magnetization $m_0$ and the density of
solitons $\bar {q}$. We present results for the case $\lambda$=0, but similar results are obtained for finite
Jahn-Teller coupling.

\begin{figure}
 \includegraphics[clip,width=9.5cm]{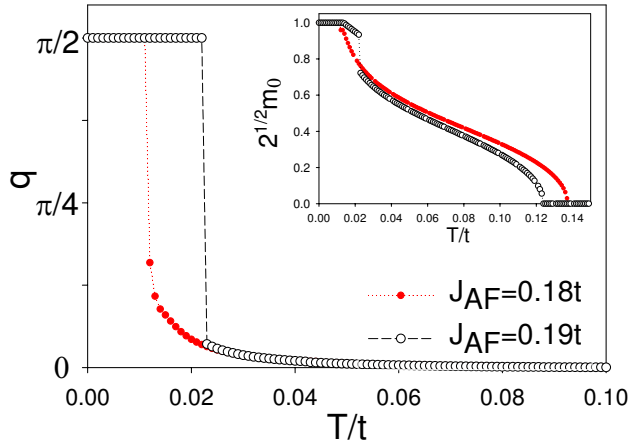}
 \caption{(Color online). Temperature  dependence of the magnetic wave vector, $q$, and of
the order parameter amplitude, $m_0$ (inset), for two values of $J_{AF}$. The  wavevector $q$=$\frac{\pi}{2}$
corresponds to the $E$-phase and $q$=$0$ to the $A$-phase. The results are obtained  by minimizing the free
energy (Eq. \ref{freeenergy}). }
 \label{Fig6}
\end{figure}

If we consider only the uniform solutions, $E$ and $A$ phases, the minimization of the free energy results in
the  phase diagrams already presented in Sec. III, Fig.\ref{Fig3}. When inhomogeneous solutions are considered,
we obtain the phase diagram shown in Fig.\ref{Fig5}. Several comments on this phase diagram are in order, (i)
There is not solitons for values of $J_{AF}$ larger than the antiferromagnetic coupling corresponding to the
Lifshitz point. The paramagnetic-$E$ phase transition is first order, with a large jump in the value of $m_S$,
and therefore in the value of $w$. In this situation, large values of $w$,  the \textit{umklapp} term is much
stronger than the elastic term and the system prefers to be commensurate with the lattice. Note than in the
paradigmatic Ising model with competing interactions\cite{bak_1980} all the transitions are second order and
incommensurate phases appear at both sides of the Lifshitz point. (ii) For small values of $J_{AF}$ the elastic
term is very strong and the solution corresponds to a dense soliton phase. For small values of $J_{AF}$ the
distance between solitons  is smaller than the cutoff and we consider that this commensurate phase is actually,
in the discrete real crystal, the ferromagnetic $A$-phase. (iii) For intermediate values of $J_{AF}$, the
competition between the elastic and the \textit{umklapp} term results in the appearance of incommensurate
solitonic phases.

In Fig.\ref{Fig5}, the shadow region indicates the incommensurate  phase. The frontier of this phase with the
ferromagnetic $A$-phase is diffuse because, as we have already discussed,  the difference between a dense
soliton phase and the ferromagnetic $A$-phase is based in a criterium on the distance between solitons. Typical
temperature dependence  of the magnetic wavevector and magnetization amplitude near the incommensurate phase  is
illustrated in Fig.\ref{Fig6}. For low temperature the system is in the commensurate $E$-phase, corresponding to
a wave vector $q$=$\frac{\pi}{2}$. At low temperatures the spins are highly polarized and that makes the
magnetic modulation too {\it rigid} to allow solitons. As temperature increases, the amplitude of the spin-wave,
$ m_0$ decreases and, at a temperature that depends on $J_{AF}$, a jump to the solitonic phase takes place.
For  values of $J_{AF}$ closer to the Lifshitz point,  the incommensurate phase appears at temperatures close to
the $A$-paramagnetic critical  temperature. In that case the amplitude of the magnetization $m_0$ is small and
hence the magnetic wavevector of the incommensurate phase is also very small. Therefore in this part of the
phase diagram the incommensurate phase is similar to the $A$-phase.

For values of $J_{AF}$ near the zero temperature $A$ to $E$ phase transition, there is  a small portion of the
phase diagram  where by decreasing the temperature the system evolves first from a paramagnetic, $q=0$, phase to
a incommensurate phase characterized by a finite $q$ and then to a ferromagnetic $A$ phase without magnetic
modulation, $q=0$.

The  magnetic phase diagram  shown in Fig.\ref{Fig5} contains the essence of the magnetic properties
experimentally observed in undoped manganites\cite{Kimura_2003}. Systems with large hopping amplitude,
($J_{AF}/t$ small) as LaMnO$_3$ have a ground state with a magnetic order of type $A$ and a relatively large
Neel temperature. The relative value of the AFM coupling increases when the ionic radius of the rare earth in
the manganite increases.  Therefore we understand  the  experimental decrease of the Neel temperature in the
series of RMnO$_3$ (R=La, Pr, Nd, Sm) as the diminution of the $A$-paramagnetic critical temperature when
$J_{AF}$ increases, see Fig.\ref{Fig3} and Fig.\ref{Fig5}. Experimentally it is observed that for large enough
ionic radius, HoMnO$_3$,  the ground state of the undoped manganite has a magnetic order of type $E$, and
present incommensurate phases when temperature increases. We claim that this situation corresponds in the phase
diagram Fig.\ref{Fig5} to values $J_{AF}/t$ in the range 0.18-0.20. Experimentally is also observed that in some
compounds as TbMnO$_3$ and GdMnO$_3$, when the temperature increases the system undergoes  two phase
transitions, first a a ferromagnetic-incommensurate transition and at higher temperatures a incommensurate
-paramagnetic transition. In the phase diagram presented in Fig.\ref{Fig5}  this behavior occurs for  values of
$J_{AF}$ near 0.18$t$.

\section{Summary}

By starting from a microscopic Hamiltonian we have derived an expression for the free energy of undoped
manganites. Using a realistic  model we have quantified the competition between the short range superexchage
antiferromagnetic interaction, and the long range double exchange ferromagnetic interaction. The competition
between these interactions results in the existence of  magnetic incommensurate phases as the  recently
experimentally observed  in undoped manganites. The incommensurate phases can be described as arrays of domain
walls separating   commensurate phases  by a distance that depends on temperature. The results presented in
Fig.\ref{Fig5} explain qualitatively the experimental results published in ref.\cite{Kimura_2003}

I

\vspace{1.truecm}

{\it Acknowledgements.} Financial support is acknowledged from Grant No MAT2002-04429-C03-01 (MCyT, Spain).


\end{document}